# Quantifying Neural Efficiency and Capacity: A Differential Equation Interpretation of Polynomial Contrasts

Jason Steffener[1,2,3,*], Karen Li[3,4], Syrina Alain[5], Johannes Frasnelli[6,7]

1 PERFORM Center, Concordia University, 7200 Sherbrooke West, Montreal, QC H4B-1R6 Canada

2 Centre de Recherche de l'Institut de Gériatrie de Montréal, 4545, Chemin Queen-Mary, Montréal, QC H3W-1W4, Canada

3 Department of Psychology, Concordia University, 7141 Sherbrooke West, Montreal, QC H4B 1R6, Canada

4 Centre for Research in Human Development, Concordia University, 7141 Sherbrooke Street West, Montréal H4B 1R6, Canada

5 Department of Anatomy, University of Quebec in Trois-Rivières, QC, Canada

6 Center for Advanced Research in Sleep Medicine, Hôpital Sacré-Coeur de Montréal, Montréal, QC, Canada

7 Research Chair UQTR Chemosensory Neuroanatomy, Department of Anatomy, University of Quebec in Trois-Rivières, QC, Canada

* Corresponding Author
Present address:
Jason Steffener, PhD
Interdisciplinary School of Health Sciences
University of Ottawa,
200 Lees, Lees Campus,
Office # E250E,
Ottawa, Ontario, CANADA
K1S 5S9

Email: Jason.Steffener@uottawa.ca

## Abstract

Task based neuroimaging tools for the study of cognitive neuroscience provide insight into understanding how the brain responds to increasing cognitive demand. Theoretical models of neural-cognitive relationships have been developed based on observations of linear and non-linear increases in brain activity. Neural efficiency and capacity are two parameters of current theoretical models. These two theoretical parameters describe the rate of increase of brain activity and the upper limits of the increases, respectively. The current work demonstrates that a quadratic model of increasing brain activity in response to the $n$-back task is a solution to a differential equation model. This reinterpretation of a standard approach to analyzing a common cognitive task provides a wealth of new insight. The results include brain wide measures of neural efficiency and capacity. The quantification of neural-cognitive relationships provides evidence to support current cognitive neuroscience theories. In addition, the methods provide a framework for understanding the neural mechanisms of working memory. This allows estimation of the effects of experimental manipulations within a conceptual research framework. The proposed methods were applied to twenty-one healthy young adults while engaging in four levels of the $n$-back task. All methods are easily applicable using standard current software packages for neuroimaging.

# 1 Introduction

A core tenet in cognitive neuroscience is the understanding of how the brain responds to increasing levels of cognitive load. Functional brain imaging provides a window into the working brain and how it responds to changing cognitive demands. Experiments demonstrate that the level of measured brain activity does indeed increase as cognitive load increases with observations including linear and nonlinear neural-cognitive relationships. Theories have developed describing these observed neural-cognitive relationships (Barulli & Stern, 2013; Grady, 2012; Rypma & Prabhakaran, 2009; Schneider-Garces et al., 2010). One basis to these theoretical models is the description of neural efficiency and neural capacity.

**Neural efficiency** describes the rate of increasing measures of brain activity as cognitive load increases. More efficient neural processing is one where the increase in brain activity per unit increase in cognitive load is smaller than a less efficient process. **Neural capacity** describes the maximal level of brain activity reached as cognitive load increases. This description of neural capacity has been expanded to describe three scenarios (Bennett, Rivera, & Rypma, 2013). The first, is when task-related brain activity increases with increasing task demands and does not reach a plateau, or capacity. This region is described as being unconstrained by capacity. A second description, is when brain activity begins to increase at low levels of demand and plateaus as demand continues to increase. This brain region is described as having a capacity independent response. The third scenario, is when a brain region demonstrates increases in brain activity with increasing task demands up until a certain level of cognitive demand. After this point, brain activity decreases with increasing cognitive load. This is described as capacity dependent. This specific concept of capacity dependence is similar to the Compensation-Related Utilization of Neural Circuits Hypothesis (CRUNCH) (Reuter-Lorenz & Lustig, 2005).

The concepts of neural efficiency and capacity have been unified into sigmoidal models of the neural-cognitive relationship (Callicott et al., 1999; Rypma, Eldreth, & Rebbechi, 2007; Schneider-Garces et al., 2010). The sigmoidal model has at minimum three parameters to describe its rate of increase and capacity limitations. This theoretical model captures many of the dynamics observed from previous studies; however, it has yet to be explicitly tested against actual brain imaging data. This lack of explicit use may be due to the model's complexity. It is a multiple parameter model requiring non-linear iterative processes for fitting to data. When combined with multi-voxel brain imaging data, such a procedure requires large amounts of computational resources and time.

The current work aims to provide a method for filling the gap in the literature between theories about neural-cognitive relationships and explicit tests of the theories. The presented approach utilizes the well-studied *n*-back working memory task and interprets results using a differential equation model. This is done by re-examining the second order polynomial model (quadratic) of brain activation with increasing cognitive load. Such a model includes intercept, linear and quadratic components when modeling neural-cognitive relationships. The innovation in the current work is that this straightforward quadratic model will be interpreted with respect to the fact that it represents a solution to a first order differential equation. The application of this polynomial regression model is therefore used with a differential equation and a "language of change" (Deboeck, Nicholson, Kouros, Little, & Garber, 2016).

Previous descriptions of brain activation increases with cognitive load (Jonides et al., 1997) may be reframed using a differential equation model and the language of change. Neural efficiency is the rate of change of brain activity as cognitive load increases. This is a measure of slope, or velocity, and the description of a first order differential equation. Neural capacity is the point in the neural-cognitive relationship where the slope stops changing, or where the differential equation equals zero.

This work uses the language of change and differential equations to model fMRI data collected from young adults while performing the *n*-back working memory task. We demonstrate that the second order polynomial linear regression model is a solution to a differential equation model and easily tested using existing software packages. Neural efficiency and neural capacity are quantified across the brain. Other measures resulting from this approach are presented, such as cognitive capacity. In summary, this technical note demonstrates the exact procedures required for fitting a current theory of neural-cognitive relationships that involves neural efficiency and capacity. The use of the well-studied *n*-back task, and easy to implement procedures, allows broad application of this approach to many research domains of cognitive performance.

## 2 Methods

### 2.1 Differential Equation Model: the language of change

The expected relationships between increasing cognitive load and measured brain activity are described using the language of change, in other words, in terms of differential equations. The possible neural-cognitive relationships are as follows. These descriptions use the parameter *'x'* to represent the level of brain activity and *'t'* to represent cognitive load. First, there may be no change in brain activity as cognitive load increases which is represented with the differential equation:

$$\frac{dx}{dt} = 0$$

This states that changes in x (dx) per unit time (dt) equal zero. Integrating this to obtain a function of the level of brain activity is:

$$x(t) = C$$

Therefore, measured brain activity is at the level C for all cognitive loads t.

Secondly, brain activity may change at a constant rate as cognitive load increases:

$$\frac{dx}{dt} = B$$

Integrating provides:

$$x(t) = Bt + C$$

Therefore, brain activity is modeled as a linear function of cognitive load with slope B and intercept C. In other words, brain activity is at the level C at the lowest level of cognitive load and then changes at a constant velocity B with increasing levels of cognitive load.

Finally, the rate of brain activity change may itself change as a linear function of cognitive load:

1)  $$\frac{dx}{dt} = 2At + B$$

Integrating this to obtain a function of brain activity provides:

2)  $$x(t) = At^2 + Bt + C$$

Therefore, brain activity is a quadratic function of cognitive load with a quadratic term A, slope B and intercept C. This differential equation can also be differentiated to obtain acceleration:

$$\frac{d^2x}{dt^2} = 2A$$

Therefore, the change in brain activation for a unit increase in cognitive load differs depending on the levels of cognitive load. Within the context of the *n*-back task, the brain activation changes from 1 to 2-back may be different from the brain activation changes from 2 to 3-back. Therefore brain activation changes its slope as cognitive load changes. In other words, brain activation accelerates or decelerates as a function of cognitive load.

This demonstrates that the second order polynomial (quadratic term) in equation 2) is the solution of the differential equation in 1). The value in this observation, is that now the tools for interpreting and analyzing differential equations may be applied to study this model of the neural-cognitive relationship. Additionally, fitting the polynomial model of brain activity with increasing cognitive load is possible with general linear modeling available in all standard fMRI analysis packages.

Therefore, this model describes brain activity with a changing velocity (slope) as cognitive load changes. This reinterpretation of the quadratic polynomial as a solution to a differential equation offers greater insight into how to interpret the parameters. To summarize, fitting the quadratic polynomial model in equation 2 to the estimated levels of brain activity from the *n*-back task provides the following. First, estimates are calculated of how the slope of brain activity changes as cognitive load changes, this is the acceleration parameter A. Secondly, estimates are calculated of the slope of brain activity as cognitive load changes, the velocity parameter B. Finally, estimates are calculated of the intercept level of brain activity, the baseline parameter C. These three parameters are calculated through estimation of specific contrasts after using a block-based model of first level fMRI brain activity. Using the language of change to describe neural-cognitive relationships facilitates the calculation of capacity and efficiency.

## 2.2 Conversion to Percent Signal Change

Although this is not a necessary step, converting brain activation maps to percent signal change units facilitates interpretation of results. This step standardizes units of brain activation allowing direct interpretation of measures within and across participants. To do this, the beta maps calculated for each level of cognitive load are divided by the estimated mean, $\beta_5$ and multiplied by 100.

$$psc\beta_i = 100 \cdot \beta_i/\beta_5, \text{ for } i = 1:4$$

## 2.3 Contrasts

The time-series data for the fMRI data is first modeled at each of the task levels as separate blocks. The next step is to calculate the contrasts appropriately so that the parameters of the quadratic equation (eqn. 2) are scaled correctly. The aim of this step is not to calculate contrasts for statistical testing. The aim is to calculate contrasts so that contrast weighted maps are correctly scaled with respect to the parameters of the quadratic equation (eqn. 2). Table 1 lists the appropriate contrast values which are obtained by taking the pseudoinverse of the unnormalized weights.

Table 1 Contrast weights for estimated model parameters.

| | Contrast Weights | | | |
|---|---|---|---|---|
| | 0-back | 1-back | 2-back | 3-back |
| C | 0.95 | 0.15 | -0.15 | 0.05 |
| B | -1.05 | 0.65 | 0.85 | -0.45 |
| A | 0.25 | -0.25 | -0.25 | 0.25 |

Notes: A, B and C are the parameters of equation 2. C: intercept, B: linear, A: quadratic

## 2.4 Cognitive Capacity

Cognitive capacity as defined here is the level of cognitive load reached when brain activity reaches its capacity level. Brain activity reaches its capacity when its first derivative (equation 1) equals zero.

Setting equation 1 to zero and solving for cognitive load gives:

3)     $Cog_{cap} = \frac{-B}{2A}$

This calculated parameter is only interpretable when brain activity reaches a maximal value which can only occur when A is less than zero, as can be seen in Figure 1, C, D, E. Appropriate scaling of the contrast weights is essential to ensure that the value calculated here is appropriate.

## 2.5 Neural Capacity

The neural capacity is the maximal brain activation value reached as cognitive load increases. This measure is again only relevant for capacity limited situations when A is less than zero. Plugging the value in 3) into equation (2) gives:

$N_{cap} = x(Cog_{cap}) = A\frac{B^2}{4A^2} - \frac{B^2}{2A} + C$

$N_{cap} = \frac{B^2}{4A} - \frac{B^2}{2A} + C$

$N_{cap} = \frac{-B^2}{4A} + C$

## 2.6 Neural Efficiency

The efficiency of the system reflects the rate at which the brain activity increases as a function of the increasing load. Therefore, neural efficiency is calculated as the velocity (slope) of brain activity between the lowest cognitive load level and the cognitive capacity. This is calculated using $N_{cap}$, C and $Cog_{cap}$:

$N_{eff} = \frac{N_{cap} - C}{-B/(2A) - 0}$

$N_{eff} = \frac{\frac{-B^2}{4A} + C - C}{\frac{-B}{2A}} = \frac{-B^2}{4A}\frac{2A}{-B}$

$N_{eff} = B/2$

Again, this calculation is only interpretable in the capacity limited situations when A is less than zero.

The three scenarios laid out by Bennett et al. (2013) for neural capacity may be interpreted here (Bennett et al., 2013). In order to differentiate neural capacity independent, unconstrained and dependent measures, the cognitive capacity term needs to be investigated further. Cognitive capacity is the cognitive load at which brain activity reaches its maximum. The definitions by Bennett et al. refer to whether the neural capacity is reached within the cognitive demands of the task. If the cognitive capacity is greater than the maximum cognitive load, then the brain region may be interpreted as having capacity independent brain activation. In the case of the current task the limit is 3.

$$\frac{-B}{2A} > 3$$

This equation may be rewritten in terms of neural efficiency as:

4)      $B/2 <- 3A$

Therefore, the neural efficiency needs to be less than negative 3 times the quadratic parameter A. If the cognitive capacity is less than the maximum cognitive load of the experiment than the brain activity in the region is interpreted as being dependent on its capacity. This describes the observation that brain activity begins to decrease past some level of cognitive load. This situation is observed by switching the comparison sign in equation 4 above.

The third scenario is when the level of brain activation continues to increase without restriction as cognitive load increases. This unconstrained capacity is observed when the A parameter is greater than zero. This interpretation of the measures within the framework laid out by Bennett et al. highlights the relationship between neural efficiency and capacity (Bennett et al., 2013).

## 2.7 Area Under the Curve

The differential equation model may be an appropriate model of the system under study, but the question remains as to the best way to determine whether there is a significant amount of task related brain activation. The current work is focused on task related increases in activation. This suggests that tests of the linear effect, parameter B, are appropriate. Simply testing for the size of B however, does not take into account the effects of the quadratic term. Testing the quadratic term alone also does not incorporate any of the linear or baseline effects. In the absence of a linear effect, a significant quadratic effect is difficult to interpret and potentially misleading.

The current work proposes the use of the area under the curve (AUC) as an appropriate assessment of task related brain activity. The AUC is the integral of the system model (equation 1) between the lowest and highest level of cognitive load. These are zero and three-back:

$$\int_0^3 (At^2 + Bt + C)dt = \frac{4}{3}t^3 + \frac{B}{2}t^2 + Ct$$

The limits, when tested are:

5)      $\frac{27A}{3} + \frac{9B}{2} + 3C = 9A + 4\frac{1}{2}B + 3C$

This can also be rewritten in terms of the contrast weights for the four levels of cognitive load allowing for parametric testing within a standard analysis package.

$$9(0.95\beta_1 + 0.15\beta_2 - 0.15\beta_3 + 0.05\beta_4) + 4.5(-1.05\beta_1 + 0.65\beta_2 + 0.85\beta_3 - 0.45\beta_4) + ...$$
$$3(0.25\beta_1 - 0.25\beta_2 - 0.25\beta_3 + 0.25\beta_4)$$

After multiplication and combining terms, this reduces to the following contrast weights:

6)    $4.575\beta_1 + 3.525\beta_2 + 1.725\beta_3 - 0.825\beta_4$

Alternatives to calculating AUC are to only calculate increases in brain activity above baseline levels. This involves removal of the 3C term in equation 5 and recalculating the contrast weights as:

$$9(0.95\beta_1 + 0.15\beta_2 - 0.15\beta_3 + 0.05\beta_4) + 4.5(-1.05\beta_1 + 0.65\beta_2 + 0.85\beta_3 - 0.45\beta_4)$$

7)    $3.825\beta_1 + 4.275\beta_2 + 2.475\beta_3 - 1.575\beta_4$

To summarize, the parameters are calculated for each of the situations plotted in Figure 1 and shown in Table 2. For comparative purposes, parameters for a model excluding the quadratic term are also included. The analysis steps are listed as a flowchart in Figure 2.

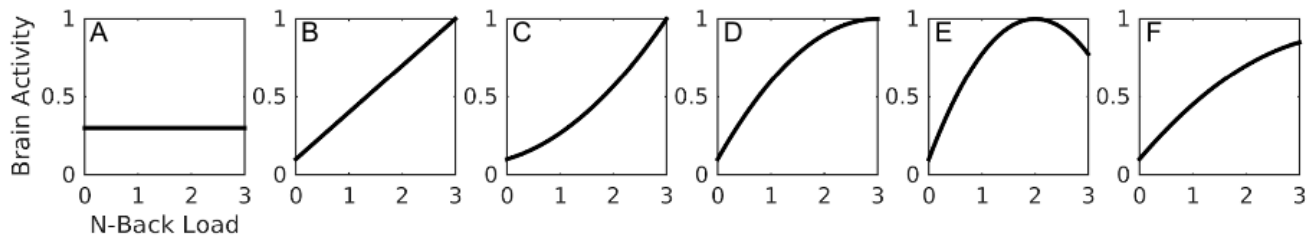

Figure 1. Six example neural-cognitive relationships. These six examples demonstrate possible neural-cognitive relationships that the quadratic model can fit. A) Constant level of neural activity. B) Linear increase in neural activity. C) Accelerating increase in neural activity also referred to as capacity unconstrained. D) Decelerating increase in neural activity, capacity independent. E) Decelerating and decreasing neural response, capacity dependent. F) Decelerating increase in neural activity, capacity independent, with lower capacity. The parameters for these six examples are in Table 2.

Table 2 Estimated parameters for simulations shown in Figure 1.

| | $x(t) = At^2 + Bt + C$ | | | | | | | | $x(t) = Bt + C$ | |
|---|---|---|---|---|---|---|---|---|---|---|
| | A | B | C | $C_{cap}$ | $N_{cap}$ | $N_{eff}$ | AUC | scAUC | B | C |
| Panel A | 0 | 0 | 0.3 | -- | -- | 0 | 0.9 | 0 | 0.0 | 0.3 |
| Panel B | 0 | 0.3 | 0.1 | -- | -- | 0.3 | 1.65 | 1.35 | 0.3 | 0.1 |
| Panel C | 0.067 | 0.1 | 0.1 | -- | -- | 0.3 | 1.35 | 1.05 | 0.3 | 0.033 |
| Panel D | -0.1 | 0.6 | 0.1 | 3.00 | 1.00 | 0.3 | 2.1 | 1.8 | 0.3 | 0.2 |
| Panel E | -0.225 | 0.9 | 0.1 | 2.00 | 1.00 | 0.45 | 2.325 | 2.025 | 0.225 | 0.325 |
| Panel F | -0.05 | 0.4 | 0.1 | 4 | 0.9 | 0.2 | 1.65 | 1.35 | 0.25 | 0.15 |

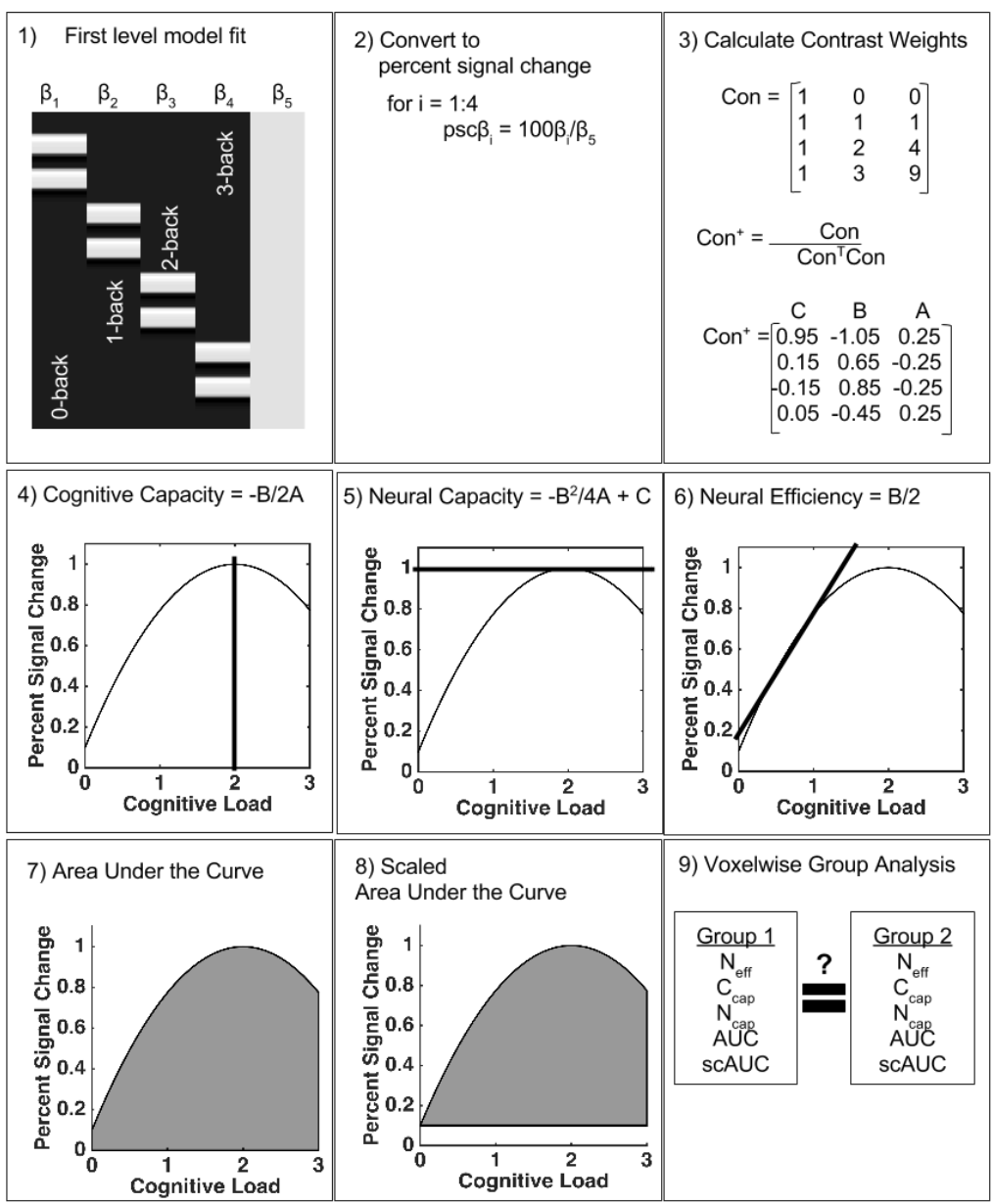

Figure 2. Flow chart of the steps for the procedures described. 1) Fit the first level model for the four levels of cognitive load and the intercept column of ones. 2) Convert the beta maps to percent signal change. 3) Calculate the contrast weights for the intercept, linear and quadratic terms. Use these contrasts to calculate the parameters A, B and C. 4) Calculate cognitive capacity. 5) Calculate neural capacity. 6) Calculate neural efficiency. 7) Calculate area under the curve. 8) Calculate scaled area under the curve. 9) Use calculated parameters for group analyses.

## 2.8 Participants

Twenty-four healthy young participants were were recruited for this study. Data from twenty-one was included in this analysis (mean (sd) age = 23.4 (3.4), 10 male, 11 female). Two participants had below chance levels of task accuracy and one had MRI data where coregistration was not possible. All participants were recruited from the Université du Québec à Trois-Rivières, Quebec, Canada. The study was approved by the Ethics Board of the Regroupement de Neuroimagerie du Québec (# CMER RNQ 15-16-10). All participants provided written informed consent.

## 2.9 Stimulus presentation

Task stimuli were back-projected onto a screen located behind the MRI bed using an LCD projector. Participants viewed the screen via a mirror system located in the head coil and, if needed, had vision corrected to normal using MR compatible glasses. Responses were made on an MRI-compatible Fiber Optic Response Pad (Current design INC.) in their right hand. The response box was connected to a laptop computer presenting the stimuli. Stimuli were presented with E-prime, Version 2.0.

## 2.10 MRI data acquisition

Images were acquired on a 3 T Siemens Tim Trio scanner with a 32 channel phased-array head coil at the "Unité de Neuroimagerie Fonctionnelle" (University of Montreal). The scanning session included an anatomical T1-weighted structural brain image obtained with an ME-MPRAGE 4-Echo sequence (176 slices, 1 mm3 voxels, TR = 2530 ms, TE = 1.64/3.5/5.36/7.22 ms, flip angle = 7°), which has a low distortion and high signal-to-noise ratio (van der Kouwe, Benner, Salat, & Fischl, 2008). Functional data were acquired with an echo planar imaging (EPI) pulse sequence (208 acquisitions, TR = 2500 ms, 41 slices, matrix size 74 × 74 voxel size 2.973 × 2.973 × 3 mm$^3$ , slice thickness: 3 mm with a 0 mm gap, TE = 20 ms, flip angle = 90°).

## 2.11 Experimental Task

The *n*-back task used letters as visual stimuli and four levels of difficulty in a fixed order (0, 1, 2, 3, 0, 1, 2, 3-back). Each of the eight task blocks involved the presentation of sequences of sixteen letters. Each letter was presented for 1.5 seconds with a 0.5 second blank between each letter. Therefore, each stimulus block was 32 seconds in duration and alternated with 24 second periods where only a crosshair was presented on the screen. The 0-back condition required participants to respond by pressing a button with the right index finger if the letter on the screen was "X" and with the right middle finger if it was not. The 1-back required a right index finger button response when the current letter matched the previously presented letter. A right middle finger response was expected when the two consecutive letters did not match. The 2-back task required matching between the current letter and that presented two letters previous. The 3-back task required matching between the current letter and that presented three letters previous.

## 2.12 Image pre-processing

All image pre-processing and statistical analyses used SPM12 (Wellcome Department of Cognitive Neurology). For each participant's EPI dataset: images were temporally shifted to correct for slice acquisition order using the first slice acquired in the TR as the reference. All EPI images were corrected for motion by realigning to the first volume of the session and the mean time-series image was calculated and written to disk using second degree B-spline interpolation. The mean EPI image was coregistered to the structural T1 image using normalized mutual information and the calculated parameters applied to all images in the time-series. The realigned and coregistered images of the time-series were written to disk using fourth degree B-spline interpolation and then smoothed with a 8x8x8mm Gaussian smoothing kernel. The resultant images were used for first-level within participant statistical modeling.

## 2.13 First Level Modeling

First-level time-series analyses used a block-based model composed of epochs separately representing each of the four cognitive loads of the $n$-back task. Each epoch was convolved with a canonical model of the hemodynamic response function supplied with SPM12. The resultant beta maps were converted to percent signal change (PSC) maps as described above. Using the PSC beta maps, five contrasts were calculated to derive the A, B and C parameters of the quadratic equation and the unscaled and scaled area under the curve values. The appropriate contrast maps were used to calculate cognitive capacity, neural capacity and neural efficiency. These calculations all used the "fslmaths" tool provided with the FSL software package (Jenkinson, Beckmann, Behrens, Woolrich, & Smith, 2012).

## 2.14 Second level modeling

Second level modeling was performed across all participants in the study. The structural image from each participant was used to calculate the individual transformations to standardized MNI space. These transformations were applied to the first level resultant images. Images were resliced to voxel dimensions of $2mm^3$ using fourth degree B-spline interpolation. Recall that all images were previously coregistered to the anatomical image. Localization of significantly high levels of brain activation used the AUC contrast from each participant in a voxel-wise single tailed one-sample $t$-test. Results in the positive direction were thresholded using false discovery rate (FDR) correction of multiple comparisons with $\alpha = 0.05$ (Genovese, Lazar, & Nichols, 2002). This corresponded to a height value of $t = 3.48$; in addition, a cluster extent threshold of 50 contiguous voxels was also used. Similar group level analyses were performed on all other participant level resultant images. The calculated beta images served as the estimated voxel-wise mean of the effects and was used for interpreting the results from the AUC contrast. All results are shown in Table 3 and overlays in Figure 3. Group mean effects over cognitive load levels for select brain locations are also shown in Figure 3.

## 2.15 Behavior

Effects of load on task performance were analyzed using a repeated measures general linear model (GLM) one-way analysis of variance with the within subjects factor of load (0-, 1-, 2- and 3-back) for both response time and accuracy. The relationships between the derived brain measures and task performance were explored. For each significant location in Table 3 the participants were grouped based on whether they demonstrated capacity constrained or unconstrained brain activation. The derived measures are applicable to capacity constrained measures, when the parameter A is less than zero. First, the two groups were compared to see if they differed in their response time or accuracy at the hardest level of task difficulty. Secondly, in participants and locations having capacity constrained responses, the derived parameters were correlated with response time and accuracy at the highest load level.

The The Rey Auditory Verbal Learning Test (RAVLT) was administered to all participants as a task

independent cognitive measure (Schmidt, 1996).

## 3 Results

Group level brain imaging results were robust when testing the area under the curve contrast. Results included many of the commonly observed regions of brain activity including: bilateral inferior, middle and superior frontal gyri, supplementary motor area, cingulate gyrus, insula, posterior parietal and occipital cortex. The measures of neural capacity are comparable across brain regions due to the initial scaling of beta maps to percent signal change. The range of percent signal change increase above baseline with increasing cognitive load ranged from 0.1% in the right middle prefrontal gyrus to 0.35% in right precentral gyrus. The neural efficiency measure indicates the estimated percent signal change per unit increase in cognitive load, e.g. going from 1 to 2 back. The estimated parameters are interrelated in the following way. Brain activity begins at the level calculated by the intercept term C. It then increases at the rate indicated by the neural efficiency measure. This increase continues until the estimated cognitive capacity for the location is reached. The value of brain activity once cognitive capacity is reached in the neural capacity. Results from the first cluster in Table 3, the insula, are used as an example: $C + C_{cap}*N_{eff} = N_{cap}$; $0.057 + 1.68*0.097 = 0.22$.

Area under the curve and scaled AUC values are also listed in Table 3. These estimates demonstrate the proportion of the increased level of brain activation above baseline that is related to working memory load related responses (sAUC) and the amount that is task related and independent of load (AUC - sAUC = 3*C). Note that Table 3 was created using the AUC contrast; therefore, all of the AUC values are significantly large. Finally, Table 3 contains capacity judgements for each listed brain region. These are defined using the criteria laid out by Bennett et al. 2013.

### 3.1 Behavior

Mean and standard deviations of response time (in seconds) for the *n*-back loads (0, 1, 2, 3) were: 0.516 (0.143), 0.564 (0.155), 0.654 (0.152), 0.700 (0.204) and for accuracy: 0.895 (0.126), 0.848 (0.196), 0.788 (0.194), 0.700 (0.174). The correlations between accuracy and response times at each load were: 0-back $r = 0.15$, $p = 0.33$; 1-back $r = -0.21$, $p = 0.35$; 2-back $r = 0.22$, $p = 0.34$; 3-back $r = 0.56$, p = 0.01. Repeated measures one-way ANOVA revealed a significant effect of load for response time ($F(3, 60) = 8.925, p < 0.001$) and for accuracy ($F(3,60) = 10.908, p < 0.001$). Mean and standard deviation of the RAVLT were: 60.0 (6.83).

For all significant locations listed in Table 3 the derived results were related to performance measures of response time and accuracy at the highest load level, 3-back, see Table 4. Participants were first split into whether they demonstrated constrained or unconstrained task-related brain activation at the location. The number of participants demonstrating each effect are listed in Table 4. Using these group splits, RT and accuracy were compared with two-sample t-tests. For most brain locations, there were more participants with constrained brain activation. Therefore, the brain activation reached a plateau either within or above the cognitive demands of the task. These results were largely inconclusive. For participants with constrained brain activity responses, the estimates of neural efficiency, cognitive capacity and neural capacity were correlated with RT, accuracy and RAVLT scores. Only a relatively few number of brain regions demonstrated significant correlations. The correlations with neural efficiency were nearly all in the positive direction. Therefore, the larger the rate of increasing brain activation, the longer their response times, the greater their accuracy and better performance on the RAVLT. This reflects the speed-accuracy tradeoff during the 3-back task. The relationships with the capacity measures were in both directions depending on the brain location.

In Table 4 the participants were split based on the capacity constrained or unconstrained brain activation in each brain region. Table 5 presents the number of brain regions having capacity constrained or unconstrained levels of brain activation for each participant. The total number of brain locations is 33. These results demonstrate that capacity constrained brain activation is more likely than unconstrained brain activation.

Table 3 Significant results from the area under the curve test

| Region | Lat | BA | x | y | z | T | k | A | B | C | $N_{eff}$ | $N_{cap}$ | $C_{cap}$ | AUC | sAUC | Cap |
|---|---|---|---|---|---|---|---|---|---|---|---|---|---|---|---|---|
| Insula | R | 47 | 40 | 24 | -4 | 8.40 | 519 | -0.058 | 0.194 | 0.057 | 0.097 | 0.221 | 1.684 | 0.527 | 0.355 | Dep |
| Inf. Frontal Oper. | R | 47 | 46 | 18 | 0 | 6.97 | -- | -0.049 | 0.163 | 0.064 | 0.082 | 0.201 | 1.676 | 0.489 | 0.296 | Dep |
| Sup. Temp. Pole | R | -- | 56 | 20 | -4 | 4.81 | -- | -0.042 | 0.135 | 0.098 | 0.068 | 0.205 | 1.593 | 0.519 | 0.226 | Dep |
| Mid. Frontal | L | 46 | -34 | 54 | 16 | 7.84 | 1081 | -0.028 | 0.135 | 0.063 | 0.067 | 0.227 | 2.439 | 0.547 | 0.357 | Dep |
| Inf. Orb. Frontal | L | 47 | -34 | 26 | -4 | 6.91 | -- | -0.034 | 0.146 | 0.079 | 0.073 | 0.234 | 2.121 | 0.586 | 0.348 | Dep |
| Insula | L | -- | -40 | 16 | 2 | 5.11 | -- | -0.021 | 0.058 | 0.074 | 0.029 | 0.114 | 1.372 | 0.293 | 0.071 | Dep |
| Supp. Motor Area | L | 32 | -2 | 14 | 46 | 7.75 | 5541 | -0.008 | 0.097 | 0.206 | 0.048 | 0.520 | 6.471 | 0.987 | 0.369 | Ind |
| Postcentral | L | 2 | -40 | -30 | 46 | 6.82 | -- | 0.010 | -0.018 | 0.171 | 0.013 | -99 | -99 | 0.525 | 0.012 | Unc |
| Postcentral | L | 3 | -46 | -24 | 48 | 6.71 | -- | 0.009 | -0.014 | 0.223 | 0.013 | -99 | -99 | 0.687 | 0.018 | Unc |
| Mid. Frontal | R | 45 | 46 | 36 | 30 | 6.56 | 1981 | -0.019 | 0.135 | 0.001 | 0.068 | 0.246 | 3.625 | 0.443 | 0.440 | Ind |
| Inf. Frontal Oper. | R | 44 | 50 | 12 | 24 | 6.31 | -- | -0.027 | 0.146 | 0.066 | 0.073 | 0.259 | 2.655 | 0.607 | 0.408 | Dep |
| Precentral | R | 6 | 40 | 6 | 48 | 6.13 | -- | -0.009 | 0.090 | 0.108 | 0.045 | 0.347 | 5.315 | 0.652 | 0.329 | Ind |
| Cerebellum | L | 37 | -36 | -52 | -30 | 6.31 | 1493 | -0.018 | 0.072 | 0.165 | 0.036 | 0.236 | 1.956 | 0.654 | 0.159 | Dep |
| Fusiform | L | 37 | -44 | -60 | -18 | 5.87 | -- | -0.027 | 0.116 | 0.108 | 0.058 | 0.231 | 2.124 | 0.600 | 0.275 | Dep |
| -- | R | -- | -34 | -52 | -40 | 5.59 | -- | 0.008 | 0.020 | 0.095 | 0.043 | -99 | -990 | 0.446 | 0.160 | Unc |
| Cerebellum | R | -- | 38 | -38 | -38 | 6.13 | 1128 | -0.016 | 0.050 | 0.170 | 0.025 | 0.208 | 1.519 | 0.587 | 0.076 | Dep |
| Inf. Temporal | R | 20 | 48 | -48 | -12 | 5.36 | -- | -0.034 | 0.134 | -0.00 | 0.067 | 0.128 | 1.948 | 0.287 | 0.293 | Dep |
| Cerebellum | R | -- | 50 | -62 | -22 | 4.89 | -- | -0.034 | 0.122 | 0.050 | 0.061 | 0.159 | 1.791 | 0.393 | 0.243 | Dep |
| Cerebellum | L | -- | -6 | -80 | -24 | 5.69 | 761 | -0.012 | 0.062 | 0.095 | 0.031 | 0.174 | 2.534 | 0.455 | 0.170 | Dep |
| Vermis | 7 | -- | 6 | -76 | -26 | 5.40 | -- | -0.008 | 0.020 | 0.180 | 0.010 | 0.193 | 1.258 | 0.560 | 0.019 | Dep |
| Cerebellum | R | -- | 8 | -74 | -34 | 4.68 | -- | 0.004 | -0.005 | 0.181 | 0.007 | -99 | -99 | 0.557 | 0.013 | Unc |
| -- | R | -- | -6 | -20 | -18 | 5.50 | 669 | 0.004 | 0.020 | 0.093 | 0.031 | -99 | -99 | 0.402 | 0.123 | Unc |
| -- | L | -- | 2 | -22 | -10 | 5.20 | -- | -0.008 | 0.081 | 0.009 | 0.041 | 0.213 | 5.007 | 0.320 | 0.293 | Ind |
| Pallidum | L | -- | -16 | 0 | 0 | 5.13 | -- | -0.008 | 0.036 | 0.052 | 0.018 | 0.093 | 2.254 | 0.246 | 0.091 | Dep |
| Sup. Parietal | R | 7 | 34 | -58 | 54 | 5.28 | 620 | -0.018 | 0.135 | 0.019 | 0.068 | 0.277 | 3.807 | 0.507 | 0.449 | Ind |
| SupraMarginal | R | 40 | 48 | -36 | 40 | 4.39 | -- | -0.007 | 0.073 | 0.056 | 0.036 | 0.257 | 5.538 | 0.434 | 0.268 | Ind |
| Inf. Parietal | R | 40 | 42 | -46 | 46 | 4.36 | -- | 0.001 | 0.050 | 0.066 | 0.054 | -99 | -99 | 0.434 | 0.236 | Unc |
| Mid. Occipital | L | 7 | -24 | -62 | 36 | 5.01 | 263 | -0.015 | 0.102 | 0.015 | 0.051 | 0.190 | 3.420 | 0.371 | 0.325 | Ind |
| Sup. Parietal | L | 7 | -24 | -62 | 46 | 4.88 | -- | -0.022 | 0.115 | 0.081 | 0.057 | 0.231 | 2.626 | 0.562 | 0.319 | Dep |
| Pallidum | R | -- | 20 | 4 | 0 | 4.64 | 55 | -0.026 | 0.110 | 0.003 | 0.055 | 0.118 | 2.107 | 0.267 | 0.259 | Dep |
| Sup. Orb. Frontal | R | -- | 30 | 66 | -6 | 4.42 | 252 | -0.009 | 0.088 | 0.037 | 0.044 | 0.251 | 4.842 | 0.427 | 0.315 | Ind |
| Mid. Orb. Frontal | R | -- | 30 | 54 | -16 | 4.27 | -- | -0.023 | 0.064 | 0.063 | 0.032 | 0.107 | 1.384 | 0.267 | 0.079 | Dep |
| Mid. Orb. Frontal | R | 10 | 40 | 62 | -2 | 4.04 | -- | 0.001 | 0.027 | 0.107 | 0.029 | -99 | -99 | 0.447 | 0.126 | Unc |

Notes: Lat: Laterality; BA: Brodmann Area where available; k: cluster size; A, B, C: parameters of equation 2, Neff: neural efficiency; Ncap: neural capacity; Ccap: cognitive capacity; AUC: area under the curve; sAUC: scaled area under the curve; Cap: capacity model; Dep: capacity dependent; Ind: capacity independent; Unc: unconstrained by capacity.

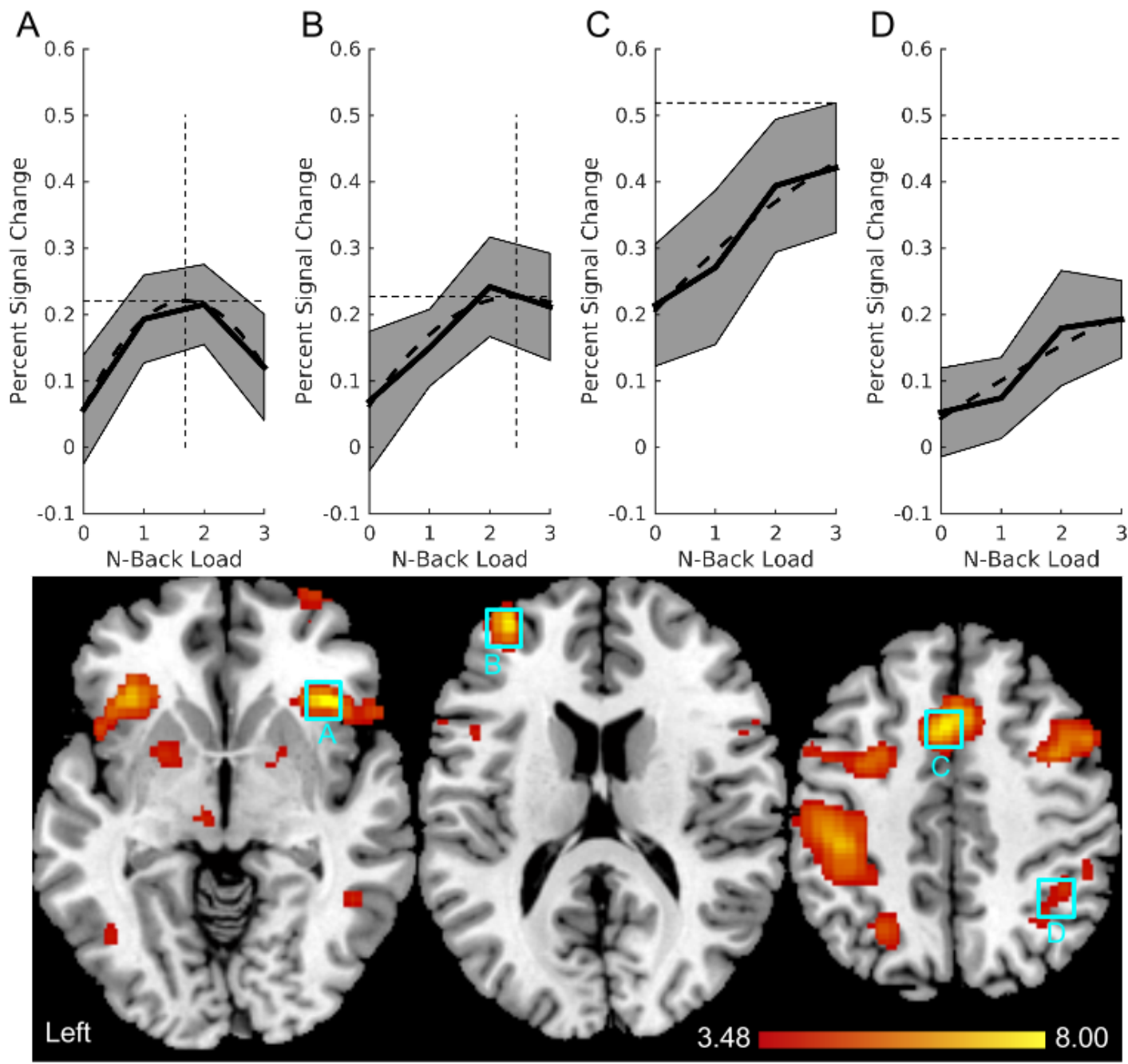

Figure 3. Brain wide results using the area under the curve contrast. This highlights brain activation in many of the commonly observed brain regions. The five highlighted brain regions demonstrate a variety of neural-cognitive relationships. A) Within the insula the brain activity was capacity dependent. Once cognitive capacity was reached, the level of brain activity decreased with further increases in cognitive load. B) Within the left middle frontal gyrus the brain activity was capacity dependent. The trajectory of brain activity reached a plateau at the highest cognitive demands of the task. C) Within the supplementary motor area brain activity was capacity independent. The neural capacity would be reached beyond the maximal level of cognitive demands. D) Within the inferior parietal region the brain activity was unconstrained by capacity. The trajectory of brain activation would not reach a capacity limit as cognitive load increased.

Table 4: relationships between derived neural measures and behavior.

| | Group Differences | | | | Correlations | | | | | | | | |
| | # Part. | | T-values | | Efficiency | | | Cog. Capacity | | | Neural Capacity | | |
| Region | Con. | Unc. | RT3 | Acc3 | RT3 | Acc3 | RAVLT | RT3 | Acc3 | RAVLT | RT3 | Acc3 | RAVLT |
|---|---|---|---|---|---|---|---|---|---|---|---|---|---|
| Insula | 16 | 5 | 0.82 | 1.38 | 0.40 | 0.42 | 0.24 | 0.19 | -0.05 | 0.01 | 0.19 | **0.52** | 0.00 |
| Inf. Frontal Oper. | 16 | 5 | **2.35** | 1.80 | 0.32 | 0.28 | 0.20 | 0.23 | -0.09 | 0.34 | -0.13 | 0.10 | 0.33 |
| Sup. Temp. Pole | 16 | 5 | 1.09 | 0.96 | **0.54** | **0.45** | 0.22 | 0.31 | -0.01 | 0.07 | 0.15 | 0.38 | 0.07 |
| Mid. Frontal | 12 | 9 | 0.09 | -0.31 | 0.30 | 0.21 | **0.48** | 0.18 | 0.30 | 0.32 | **0.50** | 0.46 | 0.32 |
| Inf. Orb. Frontal | 15 | 6 | 0.47 | 1.42 | 0.14 | **0.48** | 0.07 | 0.14 | 0.05 | 0.32 | 0.18 | 0.42 | 0.33 |
| Insula | 10 | 11 | -0.05 | -0.09 | 0.19 | 0.27 | 0.26 | 0.33 | 0.10 | 0.11 | **0.60** | **0.60** | 0.11 |
| Supp. Motor Area | 11 | 10 | 0.16 | -0.21 | 0.22 | 0.17 | 0.15 | 0.40 | 0.19 | 0.22 | 0.36 | 0.21 | 0.23 |
| Postcentral | 9 | 12 | 0.55 | -0.31 | **0.44** | 0.16 | 0.36 | -0.33 | -0.05 | **0.61** | **-0.60** | -0.37 | **0.61** |
| Postcentral | 9 | 12 | 0.94 | 0.04 | **0.44** | 0.12 | 0.37 | 0.16 | 0.39 | 0.05 | 0.14 | 0.39 | -0.11 |
| Mid. Frontal | 14 | 7 | 0.81 | -0.56 | 0.39 | 0.28 | **0.45** | -0.39 | -0.64* | 0.16 | -0.39 | **-0.65** | 0.13 |
| Inf. Frontal Oper. | 13 | 8 | -1.77 | -0.75 | 0.11 | 0.17 | 0.14 | 0.21 | 0.43 | 0.08 | 0.45 | 0.13 | 0.08 |
| Precentral | 13 | 8 | 0.03 | -0.78 | 0.19 | 0.21 | 0.32 | -0.36 | -0.10 | -0.01 | -0.10 | 0.38 | -0.01 |
| Cerebellum | 13 | 8 | 0.50 | -0.42 | -0.04 | -0.09 | -0.27 | -0.20 | -0.03 | 0.04 | 0.24 | 0.39 | 0.03 |
| Fusiform | 15 | 6 | 0.77 | -0.01 | 0.26 | 0.12 | 0.39 | -0.05 | 0.19 | -0.07 | 0.00 | 0.00 | -0.07 |
| -- | 10 | 11 | 1.82 | **2.32** | 0.06 | 0.01 | -0.22 | -0.16 | **0.60** | 0.08 | -0.03 | **-0.63** | 0.09 |
| Cerebellum | 12 | 9 | 1.54 | 1.66 | -0.15 | 0.03 | -0.21 | -0.04 | -0.34 | 0.07 | -0.05 | 0.25 | 0.01 |
| Inf. Temporal | 14 | 7 | -1.39 | 0.12 | 0.24 | 0.31 | 0.32 | **0.67** | 0.30 | 0.02 | 0.33 | 0.10 | 0.02 |
| Cerebellum | 12 | 9 | -0.69 | -1.65 | 0.27 | -0.02 | **0.49** | 0.19 | 0.04 | 0.15 | 0.31 | 0.10 | 0.15 |
| Cerebellum | 12 | 9 | 0.53 | 0.61 | 0.18 | 0.07 | 0.14 | 0.39 | 0.18 | 0.18 | 0.16 | 0.05 | 0.18 |
| Vermis | 12 | 9 | -0.77 | -0.31 | 0.05 | 0.09 | 0.10 | 0.00 | -0.06 | 0.06 | -0.06 | 0.25 | 0.07 |
| Cerebellum | 11 | 10 | 0.31 | 0.71 | 0.05 | 0.25 | 0.30 | -0.06 | -0.34 | 0.09 | -0.42 | 0.08 | 0.10 |
| -- | 10 | 11 | -0.05 | -0.09 | -0.10 | -0.20 | -0.04 | **0.58** | 0.33 | 0.11 | 0.45 | 0.28 | 0.11 |
| -- | 9 | 12 | 0.46 | -0.31 | 0.34 | 0.08 | 0.23 | 0.17 | -0.25 | 0.01 | 0.25 | 0.16 | 0.01 |
| Pallidum | 10 | 11 | -0.05 | -0.71 | 0.28 | 0.14 | 0.42 | -0.03 | -0.43 | 0.31 | 0.11 | -0.08 | 0.31 |
| Sup. Parietal | 13 | 8 | -1.26 | -0.73 | 0.05 | 0.05 | -0.16 | 0.21 | 0.37 | 0.02 | -0.25 | -0.11 | 0.03 |
| SupraMarginal | 11 | 10 | 0.77 | -0.23 | 0.04 | 0.02 | 0.36 | -0.24 | 0.10 | 0.05 | -0.34 | -0.24 | 0.05 |
| Inf. Parietal | 12 | 9 | -1.34 | -1.30 | 0.00 | 0.05 | 0.24 | -0.16 | 0.12 | -0.02 | -0.10 | 0.30 | -0.01 |
| Mid. Occipital | 12 | 9 | -1.44 | -0.40 | -0.18 | 0.25 | 0.06 | -0.33 | -0.21 | -0.11 | -0.29 | -0.22 | -0.11 |
| Sup. Parietal | 15 | 6 | -1.81 | -1.07 | -0.04 | 0.11 | 0.11 | -0.35 | -0.20 | 0.01 | -0.23 | -0.16 | 0.01 |
| Pallidum | 11 | 10 | 1.13 | -0.21 | 0.28 | 0.06 | 0.32 | 0.07 | -0.03 | 0.21 | -0.13 | -0.13 | 0.21 |
| Sup. Orb. Frontal | 11 | 10 | 1.64 | 0.09 | 0.29 | -0.01 | 0.21 | -0.55 | -0.11 | 0.05 | 0.07 | 0.02 | 0.05 |
| Mid. Orb. Frontal | 13 | 8 | 1.05 | -0.42 | 0.25 | 0.04 | 0.00 | 0.29 | 0.68 | 0.30 | -0.19 | -0.05 | 0.30 |
| Mid. Orb. Frontal | 11 | 10 | -0.82 | -1.52 | 0.22 | 0.00 | 0.25 | 0.40 | 0.27 | -0.22 | **0.57** | 0.34 | -0.22 |

*Notes*: Lat: Laterality, BA: Brodmann Area where available, k: cluster size. *: there is a single outlier driving this result. Correlations need to be above 0.43 to be significant at α < 0.05 with this sample size, significant correlations are in bold.

Table 5. Proportion of voxels per participant demonstrating capacity constraints.

| Participant ID | Capacity Constrained | | Capacity Unconstrained |
|---|---|---|---|
| | Dependent | Independent | |
| 1 | 73 | 12 | 15 |
| 2 | 67 | 6 | 27 |
| 3 | 45 | 18 | 36 |
| 4 | 79 | 12 | 9 |
| 5 | 0 | 0 | 100 |
| 6 | 42 | 3 | 55 |
| 7 | 12 | 0 | 88 |
| 8 | 33 | 3 | 64 |
| 9 | 58 | 6 | 36 |
| 10 | 39 | 18 | 42 |
| 11 | 33 | 3 | 64 |
| 12 | 36 | 0 | 64 |
| 13 | 76 | 12 | 12 |
| 14 | 88 | 3 | 9 |
| 15 | 30 | 24 | 45 |
| 16 | 61 | 0 | 39 |
| 17 | 24 | 12 | 64 |
| 18 | 36 | 12 | 52 |
| 19 | 39 | 15 | 45 |
| 20 | 67 | 24 | 9 |
| 21 | 91 | 6 | 3 |

*Note*: These proportions are for the 33 voxel locations listed in Table 3.

## 4 Discussion

This works presents an analysis and interpretation for the *n*-back task with four levels of difficulty. The research question of 'Does brain activity increase as a function of cognitive load?' was reframed into a differential equation model. Differential equations are used to describe systems and refer to changes in one variable in terms of another. Using this language of change, the research question is restated as: 'Is the rate at which brain activity changes with increasing cognitive load constant, or does it change?' The two research questions may be tested with identical regression models; however, they differ in their estimation of cross-load assessments. The benefit of rephrasing the research question in terms of differential equations is that it opens up a host of analytic and interpretative approaches that facilitate the understanding of neural-cognitive relationships.

Differential equations are set to zero and solved to determine when the functions reach maxima and minima and what the values of the function are at these values. In terms of the current experiment, these measures are as follows. The maximal value of brain activation reached with increasing cognitive load is the point when the derivative of the model (equation 1) equals zero. This is the neural capacity of the system. As discussed by Bennett et al. (2013) brain regions may be capacity dependent, capacity independent or unconstrained by capacity (Bennett et al., 2013). Using this definition, the estimated parameters can be used to make these distinctions across the entire brain.

It is also possible to calculate the value of cognitive load when the neural capacity is reached. We term this the cognitive capacity of a brain region. This measure is novel and may have its own interesting effects when linking brain and behavioral measures. It is also useful for calculating the neural efficiency of a brain region. Neural efficiency is the rate at which brain activity increases as cognitive load increases (Barulli & Stern, 2013; Stern, 2009). This is calculated as the slope of brain activity midway between the lowest level of cognitive load and the measured cognitive capacity of a brain region.

In addition to these measures of efficiency and capacity, the proposed analysis framework provides measures assessing the amount brain activity increased as cognitive load increased. This is important for identifying which brain regions respond to increases in cognitive load. The most straightforward approach is to simply test whether the linear term differs from zero. This however does not incorporate any of the neural-cognitive dynamics that are captured with the quadratic term. As seen in Table 2, it may also over or underestimate true effects in the data. To incorporate both linear and quadratic effects, measures of the area under the curve (AUC) were tested. Calculating area under the curve within the framework of differential equations is done by integrating the function between two limits. In the current case, it is integrating brain activity as a function of cognitive load, equation 1, between load levels 0 and 3-back. This measure fully captures the intercept, linear and quadratic features of the model into a single number.

Two key points must be stressed for this analysis. First, the contrast weights for quantifying the intercept, linear and quadratic effects must calculate the parameters of the quadratic function, see Table 1. Alternate sets of contrasts weights may correctly quantifying whether linear and quadratic effects are significantly present in the data. The current approach however, is interested in calculating the parameters of an equation. This is analogous to fitting a polynomial to data. While alternate contrast weights may provide similar significance test results, they do not calculate the actual model parameters.

The current results demonstrate significantly large task related increases in brain activity in brain regions typically reported with the *n*-back task (Owen, McMillan, Laird, & Bullmore, 2005). Inspecting the AUC and AUCsc results demonstrate that some brain areas may have a consistently high level of brain activity for all load levels. Scaling the AUC relative to baseline levels (AUCsc) provides a better indicator of load related increases in brain activity. As an example, brain activity in left postcentral gyrus has large AUC values: 0.52 to 0.69. Once the values are scaled the values drop to 0.01 and 0.02, respectively. This demonstrates that in this brain region there was a load independent increase in brain activation. This is a logical finding and interpretation due to the right handed buttons responses required for all levels of task load for this experiment. One advantage to calculating these measures within the differential equation framework is that the two AUC measures may be calculated using specific contrast weights within first-level modeling of the fMRI data, equations 6 and 7. This facilitates the use of the measure within existing brain imaging analysis packages.

The relationships between derived measures and behavior demonstrated sparsely significant findings;

however, nothing robust and consistent. These results are interesting; however, they must be interpreted with caution as no control for type I error was made. These results may be due to the healthy young participants used. In previous work with older adults, brain-behavior relationships were only found in older adults (Zarahn, Rakitin, Abela, Flynn, & Stern, 2006).

The current analysis did allow participants and brain regions to be investigated as to whether the brain activity response to increasing cognitive load was constrained by capacity (Bennett et al., 2013). The current approach therefore provides a quantification of this hypothetical model. Although there was a preponderance of capacity constrained results, brain regions did not demonstrate a consistently constrained or unconstrained response to increasing load across participants, see Table 5. When collapsing across brain regions, there was again a similar split with only a couple of exceptions. Two participants demonstrated mostly unconstrained task related brain activation across regions. This suggests that these participants may be able to meet the functional demands of a much higher level of cognitive load than others. Further investigation into individual differences in lifetime exposures and neuropsychological assessment may shed light on these variations.

The current approach used a quadratic polynomial linear regression model of brain activation in response to four levels of cognitive load during an $n$-back task. Theoretical descriptions in the literature often describe the relationships between task demands and brain activation as sigmoidal (Callicott et al., 1999; Rypma et al., 2007; Schneider-Garces et al., 2010). These theoretical relationships may be modeled at each voxel with sigmoidal functions such as the Weibull distribution (Trögl & Benediktová, 2011; Weibull, 1951). This class of functions are more sophisticated and capture more of the subtle hypothesized inflections in brain-behavior relationships. A drawback with sigmoidal models is that their sophistication requires a higher number of cognitive loads and nonlinear parameter estimation. Nonlinear parameter estimation is an iterative method having a high computational burden and subject to nonconvergence errors. The advantage of the presented approach is that it may be implemented in current software packages using linear regression without any computationally intensive nonlinear parameter estimation.

The current work provides an approach to complement our understanding of the neural mechanisms of working memory capacity (Cowan 2010; Miller and Buschman 2015). This approach provides explicit measures of neural capacity and efficiency which may be linked to behavioral measures. These may be $n$-back task performance or other behavioral measures of working memory capacity. As an example, integration of the proposed methods and working memory span would provide insight into the neural underpinnings of memory spans. This will be a future avenue of investigation.

Future directions may also directly follow approaches of more formal multilevel nonlinear modeling of working memory (Oberauer and Kliegl 2006). The approach of these authors is possible with brain imaging data by incorporating the proposed methods with recent statistical software advances (Chen et al. 2014; Chen et al. 2013). This would provide a descriptive model of the neural mechanisms of working memory for estimating the effects of experimental manipulations.

The term "neural efficiency" has received some recent critique (Poldrack, 2015/2). Its use here is to describe a parameter of current models of the cognitive neuroscience of aging. The current approach provides contextual interpretation for cognitive load related increases and decreases in measures of task-related brain activation. Future directions may refine the proposed methods through integration of PET based regional markers of energy usage and control of time on task effects (Grinband, Wager, Lindquist, Ferrera, & Hirsch, 2008; Poldrack, 2015/2).

Having a descriptive model of neural-cognitive relationships provides a framework for testing a variety of questions. Within the context of healthy aging one question is "Does aging affect neural efficiency or capacity?" This may then be followed up with tests of whether variations in neural efficiency and capacity are related to measures brain structure or cerebrovasculature (Steffener & Stern, 2012). Measures of neural efficiency and capacity provide a link to the physiology of brain activation and the effects of aging (D'Esposito, Deouell, & Gazzaley, 2003) beyond measures of statistical amplitude and spatial extents of task-related activation.

This work presents a method for explicit quantification of neural capacity and neural efficiency within a differential equation framework. This allows for the quantification of neural efficiency and neural capacity at the voxel-wise level. Future directions will be to compare young and old adults to explore how aging affects neural efficiency and capacity. Additional work, will test how indices of cognitive reserve, such as education, physical activity and IQ, are related to different levels of neural efficiency, neural capacity or cognitive capacity.

## 5 Conclusions

A method was presented to quantify the neural efficiency and capacity parameters of the sigmoidal neural-cognitive model. The presented methods may be completely implemented using existing standard task-based brain imaging software packages (e.g. SPM, FSL, AFNI). Data from twenty-one healthy young adults demonstrate results from using these methods. Results highlight commonly identifying regions involved in working memory paradigms. In addition to quantifying a current theoretical model, a novel interpretation framework is presented using a "language of change." This is possible through the recognition of the quadratic polynomial model as a solution to a differential equation (DE). This DE is a model where the rate of increasing brain activity changes as a function of cognitive load. The concept of a changing rate is implicit in sigmoidal models. The present work however represents one of the first explicit tests of this model and brain-wide quantification of its parameters. The use of DE models of brain measures may be expanded to model volumetric changes across the lifespan; furthermore, more sophisticated neural-cognitive relationships are also possible.

## 6 Acknowledgements

Thank you to Carollyn Hurst and Andre Cyr at the UNF for MRI scanning and the following research assistant for making data collection possible: Noémie Mercier, Joëlle Robert, Lucie C. Frenette, Florence Gingras-Lessard, Sherezada Ochoa.

## 7 Funding Sources

Quebec Bio-Imaging Network (QBIN) awarded to JF.

## 8 Conflict of Interests

The authors declare no conflicts of interests.